\documentclass[twocolumn,showpacs,preprintnumbers,amsmath,amssymb]{revtex4}
\usepackage{graphicx}
\usepackage{dcolumn}
\usepackage{bm}
\usepackage{here}
\usepackage{color}

\begin{document}

\title{Universal Irreversibility of Normal Quantum  Diffusion}
\author{Hiroaki S. Yamada}
\email{hyamada@uranus.dti.ne.jp}
\affiliation{Yamada Physics Research Laboratory, Aoyama 5-7-14-205, Niigata 950-2002, Japan}
\author{Kensuke S. Ikeda}
\email{ahoo@ike-dyn.ritsumei.ac.jp}
\affiliation{Department of Physics, Ritsumeikan University
Noji-higashi 1-1-1, Kusatsu 525, Japan}

\date{\today}
\begin{abstract}
Time-reversibility measured by the deviation of
the perturbed time-reversed motion from the unperturbed
one is examined for normal quantum diffusion 
exhibited by four classes of quantum maps with contrastive
physical nature. Irrespective of the systems, there exist a 
universal minimal quantum threshold above which the system 
completely loses the past memory, and the time-reversed 
dynamics as well as the time-reversal characteristics 
asymptotically trace universal curves independent of 
the details of the systems.
\end{abstract}

\pacs{05.45.Mt,03.65.-w,05.30.-d}

\maketitle


\def\etath{\eta_{th}}
\def\irrev{{\mathcal R}}
\def\e{{\rm e}}
\def\noise{n}
{\it Introduction: }
Historically, the irreversible time evolution was 
considered as characteristics peculiar to macroscopic 
system composed of many degrees of freedom.  However, 
since the rediscovery of chaos in systems with a small 
number of degrees of freedom (SNDF), 
chaos, which exists in non-integrable dynamical systems, 
has been considered as a more generic origin 
of time-irreversibility and dissipation \cite{prigogine81}. 

On the other hand, quantum systems has no 
counterpart of classical orbits, and chaos in the sense of classical 
mechanics does not exist. 
However, in quantum systems which is chaotic in the classical limit, 
apparently irreversible phenomena such as normal diffusion \cite{chirikov81,adachi88}, 
stationary energy absorption \cite{ikeda93} 
and so on do occur although 
the number of degrees of freedom is small. 
Moreover, even in quantum systems with no classical limit, 
normal diffusion can be realized, as is typically exemplified
by more than one-dimensional disordered systems \cite{lifshiz88}.  
However, very few works has been done for 
the direct characterization of time-irreversibility in 
apparently irreversible quantum systems. 

The purpose of this letter is to explore the time-irreversibility of
quantum dynamics exhibiting normal diffusion, which is characterized
by the stationary linear increase of mean square displacement (MSD),  
$M(t) =  \sum_{x}P(x,t)x^2 = Dt$. (The diffusion space depends on the
representation.) It is quite natural to quantify the time 
irreversibility of normally diffusing systems on the basis of MSD.

We quantitatively characterize the instability underlying 
quantum dynamics in terms of the sensitivity of time-reversed 
dynamics to the perturbation applied at a reversal time, 
which is refered as {\it time-reversal test}. 
{\it Method:}  
The time-reversal test consists of the following
three processes. First the initial point-wise localized state evolves 
in forward by operating the evolution operator $U^t$ until $t=T$. At 
the reversal time $t=T$, a perturbation $\hat{P}(\eta)$ is applied, 
and finally the perturbed state is evolved in backward by operating 
the time-reversed evolution operator $U^{-T}$. With the use of MSD, 
the relative irreversibility 
\begin{eqnarray}
\label{clsR1}
 \irrev(\eta)=\frac{|M_\eta(2T)- M_0(2T)|}{M_0(T)},
\end{eqnarray}
is defined as a function of the perturbation strength $\eta$. It 
is used as a measure of the sensitivity of the quantum dynamics to the 
external perturbation, which is refered as the {\it time-reversal
characteristics} \cite{ikeda96}. 
We would like to use $\irrev$ rather than the "fidelity", which has been 
frequently used in literatures \cite{benenti02,prosen02} since \cite{peres84},
as the measure of quantum irreversibility, because $\irrev$ allows 
an immediate comparison between quantum irreversibility and 
its classical counterpart. 

As the perturbation $\hat{P}$ applied 
at $T$, we mainly use the $\eta-$shift operator, which shifts 
the wavepacket by $\eta$ in the $y-$space canonically conjugate 
to the diffusion space $x$,  
\begin{eqnarray}
 \hat{P_y}(\eta)=\exp \{i\eta \hat{x}/\hbar\}=\exp\{\eta \partial/\partial y\}.
\end{eqnarray}
We call it the "perpendicular shift". There is also the 
"parallel shift" $\hat{P_x}(\eta)=\exp \{i\eta \hat{y}/\hbar \}$
which shifts the wavefunction in the diffusion space $x$ by
 $\eta$, but we use in this letter only "perpendicular shift"
on a reason mentioned later.
This method is a powerful tool when we measure the instability of quantum
dynamics which have no counterpart of the classical orbital instability 
\cite{ikeda96}.

{\it Models:}
As the model system for which the time-reversal test is examined, 
we use four kinds of quantum maps with quite different nature.
The models are described by the following common form of unitary 
operator, 
\def\hatp{\hat{p}}
\def\hatq{\hat{q}}
\def\hatU{\hat{U}}
\begin{eqnarray}
\label{unitary}
   \hatU  = \e^{-i\frac{H_0(\hatp)}{2\hbar}}\e^{-i\frac{V(\hatq)}{\hbar}}\e^{-i\frac{H_0(p)}{2\hbar}}, 
\end{eqnarray}
where $H_0(\hatp)$ and $V(\hatq)$ represent kinetic 
energy and potential energy, respectively. Here $\hatp$ and 
$\hatq$ are momentum and position operators, respectively.
The first example is 
standard map (SM), which is given by $H_0=\frac{p^2}{2}, V(q)=K \cos q$, 
is a typical deterministic quantum map whose classical counterpart
is chaotic if $K$ is large enough and shows a nice quantum diffusion 
in the $p$ space 
if $K\ll 1$ and $ \hbar \ll 1$. 
The second example is the perturbed Anderson 
map with $H_0(p)= \cos (\frac{p}{\hbar})$ and 
$V(q,t)=v_q \{1+  \sum_{i=1}^M \epsilon_i \cos{\omega_i t} \}$ \cite{yamada04}. 
It is a quantum map version
of Anderson model defined on the discretized lattice $q \in {\bf Z}$.
On-site potential $v_q$ taking random value uniformly distributed 
over the range $[-W, W]$ leads to the localization of wavepacket,
but the quasi-periodic perturbation of
the strength $\epsilon$ with the $M$ incommensurate frequencies 
destroys the Anderson localization, resulting into 
a well-behaved diffusion in the $q$ space
if $M\ge 2$ and $\epsilon \geq O(1)$.
We call this model the perturbed Anderson map (PAM). 
In the following numerical calculation, we take 
$W=1.0$ or $0.5$ and $\epsilon_i=\frac{\epsilon}{\surd M}$, for simplicity, 
and take incommensurate numbers of $\omega_i \sim O(1)$ as the frequency set.

The first and the second examples are both intrinsic dynamical
systems which contains no stochastic force.  We furthermore 
examine stochastically-perturbed quantum maps as another
prototypes showing noise-induced normal diffusion. 
The third example is the stochastic standard map (SSM) 
$\hat{U}=\e^{-ip^2/2\hbar} \e^{-i \epsilon \noise_{t} \sin q/\hbar} 
\e^{-ip^2/2\hbar}$ parametrically driven by the temporal 
noise $\noise_{t}$ \cite{yamada10}.
The final example is a quantum map version of Haken-Strobal 
model $\hat{U}=\e^{-i\cos(\hatp/\hbar)/2\hbar}
\e^{-i\noise_{qt}/\hbar}\e^{-i\cos(\hatp/\hbar)/2\hbar}$ driven
by spatio-temporal noise \cite{haken73}, which 
we call Haken map (HM). In both models the applied noise 
satisfy the sample average 
$<\noise_{t}\noise_{t'}>=\epsilon_n^2\delta_{tt'}$ for SSM and  
$<\noise_{q,t}\noise_{q',t'}>=\epsilon_n^2\delta_{qq'}\delta_{tt'}$ for HM, 
where $\epsilon_n$ is strength of the noise. 

In the limit $\hbar\to 0$, both SM and SSM have classical
counterparts whose orbit is exponentially unstable \cite{footnote}.
  However, PAM and HM have not their 
classical counterparts, since the transfer 
operator $i\cos(\hatp/\hbar)/2$ has no classical limit.
We may expect that quantum normal diffusion systems
having a classical limit may somehow mimic the 
time-irreversibility of the classical counterpart in the 
limit $\hbar \to 0$. But we have no reference 
based on the classical theory for PAM and HM, and so
we are particularly interested in the time-irreversibility for 
the latter class of quantum normal diffusion.

{\it Result:}
Figure 1(a) and (b) are typical results of time-reversal test examined for 
normally diffusing state of SM and PAM. For relatively strong
perturbation strength $\eta$ at the reversed time $T$, the time-reversed 
dynamics follow the unperturbed time-reversed dynamics for only a short
period, and it sooner recovers the normal diffusion at the same
diffusion constant $D$ of the forward diffusion (this is due to 
the time-reversal symmetry of our system). 
\begin{figure}[!ht]
\begin{center}
\includegraphics[width=7.5cm]{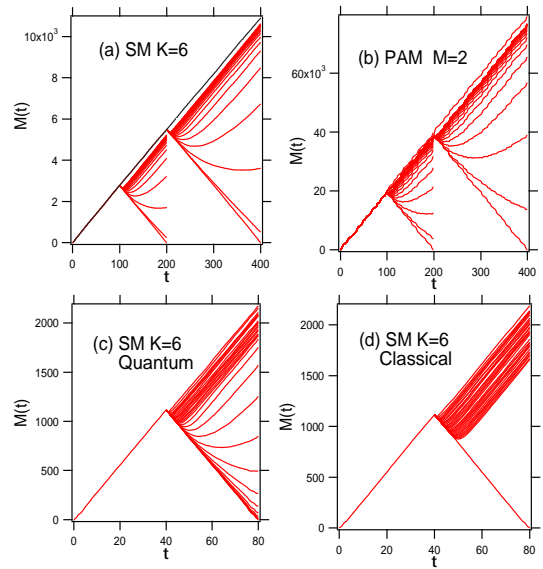}
\caption{(Color online) 
Time-reversal tests of SM and PAM with self-induced  
normal diffusion.
(a) SM with $K=6$, $\hbar=\frac{2\pi 121}{2^{17}}$.
(b) PAM with $M=2, \epsilon=1.0$, $\hbar=0.25$.
(c) Short reversed-time version of (a).
(d) Classical SM with $K=6$ corresponding to (c).
\label{fig1}}
\end{center}
\end{figure}
This fact allows us to interpret that the recovered diffusion is delayed 
from the forward diffusion $M(t)=Dt$ with the delay time $\tau_d$, namely 
asymptotically $M_\eta(t)=D(t-\tau_d)$ for $t \gg T+\tau_d$. 
However, as $\eta$ decreases, there appears a strong tendency of
following the unperturbed time-reversed dynamics. 
Such a nature is quite different from
the classical time-irreversibility typically observed for the
classical counterparts of SM, shown in Fig.1(d)

Fig.1(d) and (c) compares the 
time-reversed dynamics of classical and quantum SM in chaotic
regime. In Fig.1(d) the perturbation strength 
$\eta$ decreases geometrically, then the delay $\tau_d$ increase 
in proportion to $\eta$, which means that  $\tau_d \propto \log\eta$. 
The $\log\eta$ dependence of the delay time is the result of 
the exponential instability inherent in classical chaotic dynamics: 
the perturbed time-reversed orbit separates from the unperturbed 
one exponentially, which makes the deviation grow as   
$\Delta M_\eta(T,\tau) \equiv M_\eta(T+\tau)-M_0(T+\tau)\sim \eta\e^{\lambda \tau}$,  
where $\lambda$ is the Lyapunov exponent. The exponential deviation
changes into the delayed diffusion after the nonlinear saturation
of exponential instability $\eta\e^{\lambda \tau_d} \sim O(1)(=C)$,  
which gives $\tau_d(\eta)=\frac{\log(C/\eta)}{\lambda}$.
The classical relative irreversibility (\ref{clsR1}) is thus 
\begin{eqnarray}
\label{clsR2}
\irrev_{cl} \sim 2-\frac{\tau_d(\eta)}{T} = 2-\frac{\log C/\eta}{\lambda T}.
\end{eqnarray}
which agrees with the classical numerical result of Fig.2(a).
It means that we have to make $\eta$ exponentially as 
small as $\eta \sim C \e^{-\lambda T}$ in order to control 
the system to restore the time-reversibility. 
Numerically observed quantum $\irrev$ for various $T$, 
is presented in Fig.2. Here $\eta$ is scaled by a fundamental 
unit $\eta_{th}$, which will be discussed later in detail.  
It clearly follows the classical $\log\eta/T$ dependence of 
Eq.(\ref{clsR2}) in a relatively large regime of $\eta$,
\begin{figure}[htbp]
\begin{center}
\includegraphics[height=4.5cm]{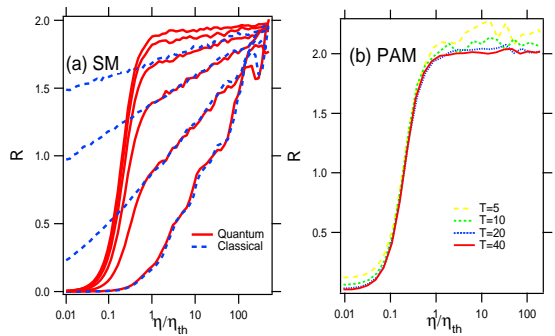}
\caption{\label{fig-SM-Reta1}
(Color online)
Time-reversal characteristics of (a) quantum (solid) and classical(broken) SM and
(b) PAM as functions of the scaled perturbation strength $\eta/\eta_{th}$ 
at several reversal  times $T=5,10,20,40,80$ from below.
$K=6$, $\hbar=\frac{2\pi 121}{2^{17}}$ for SM and $W=1.0, M=3,\epsilon=0.5$
 for PAM.
The classical time-reversal characteristics are well described by Eq.(\ref{clsR2}).
}

\end{center}
\end{figure}
but a striking difference between quantum and classical 
appears in the low $\eta$-regime: there exists a threshold 
$\eta_{th}$ below which the quantum reversed motion, 
differing entirely from the classical one, restores the 
time-irreversibility and the
relative irreversibility approaches promptly to 0.

The presence of the threshold is a direct manifestation 
of quantum uncertainly in the quantum time-irreversibility.
Suppose that the wavepacket diffuses to cover the range of $x$
width $\Delta x(T)=\sqrt{M(T)}$ at the reversal time. 
So the perpendicular perturbation (2) shifting the 
quantum state in the $y(=q)$ space by $\eta$ sweeps 
the phase space over the area $A=\eta \Delta x(T)$.
The shifted quantum state is recognized as
classically distinguishable one
from the original state if a $\eta$ is large enough such that
the sweeped area contains more than one quantum state, 
namely $A/h>1$, which defines the 
{\it least quantum perturbation unit} (LQPU),  
\begin{eqnarray}
\label{LQPQ}
   \eta_{th}=\frac{2\pi \hbar}{\Delta X}=\frac{2\pi \hbar}{\sqrt{M(T)}}
\end{eqnarray}
as the threshold perturbation strength. If $\eta>\eta_{th}$,   
the orbit from the shifted state separates from the orbit from 
the original state in the classical mechanical way. 
In Fig.2 $\irrev$ is displayed as functions of $\eta$ 
scaled by the LQPU, and so the quantum threshold is the common value $1$ in the unit.
 Hereafter, we refer the regions $\eta/\eta_{th}<1$ and
$\eta/\eta_{th}>1$ as {\it quantum region}  and
{\it post quantum region}, respectively \cite{footnote2}. 
The SSM, which has the classical counterpart, shows quite similar
behavior as in Fig.2(a).

A typical example of time-reversal
characteristics in systems without the classical counterpart
is shown in Fig.2(b). Even in this case we can recognize the
quantum and the post-quantum regions similarly to Fig.2(a), but we 
can not observe a significant $T$ dependence of the characteristics, 
and convergence to the asymptotic characteristics occurs much more
rapidly than SM and SMS. In spite of the difference in the way of
convergence, the asymptotic limit exhibits a remarkable universality
irrespective of the kind of the system independent on whether the 
system has the classical unstable limit or not, and whether 
the diffusion is due to the stochastic perturbation or not.


\begin{figure}[htbp]
\begin{center}
\includegraphics[width=6cm]{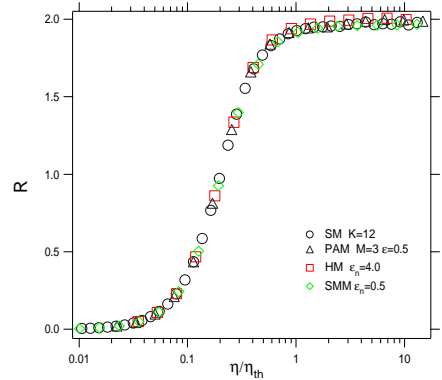}
\caption{ \label{fig3} 
(Color online)
Time-reversal characteristics as a function of the scaled perturbation strength 
for some quantum maps which exhibit normal diffusion, i.e. 
SM with $K=4.2, 12$ at $T=800$, 
PAM with $M=3, \epsilon=0.5,0.8$ at $T=200$, 
HM with $\epsilon_n=4.0$ at $T=500$, 
and SSM with $\epsilon_n=0.1$ at $T=3200$. 
 }
\end{center}
\end{figure}
Figure 3 depicts the relative irreversibility 
$\irrev$ as a function of the scaled perturbation strength for the 
four kinds of models, which is obtained in the large limit of $T$. 
All the plots are on a common universal curve, and so
the irrevrsibility $\irrev$ reaches to 2 in the same way as
$\eta/\eta_{th}$ exceeds $1$, which means that the system completely
lose the memory and reset to the stationary diffusion beyond
$\eta_{th}$ in a universal way. It is the significance of the universal 
quantum threshold $\etath$ as the LQPU. 

It seems that such a class of universality is not only a feature
of the irreversibility defined at the returning time $t=2T$
but also is a general property of the time-reversed dynamics itself.  
Let us return to the separation of the perturbed time-reversed 
process from the unperturbed one, which is represented by the 
difference $\Delta M_\eta(T,\tau)=M_\eta(\tau+T)-M_0(\tau+T)$.  
In Fig.4, the scaled separation $\Delta M_\eta(T,\tau)/M_0(T)$ 
is shown as a function of the scaled time $\tau/T$ for
SM and PAM. The observed scaled separation $\Delta M_\eta(T,\tau)/M_0(T)$ 
vs scaled time $\tau/T$  is on the common curve independent of the system
in the limit $T \to \infty$ if the scaled perturbation strength  $\eta/\etath$ 
is the same.
\begin{figure}[!ht]
\begin{center}
\includegraphics[width=6cm]{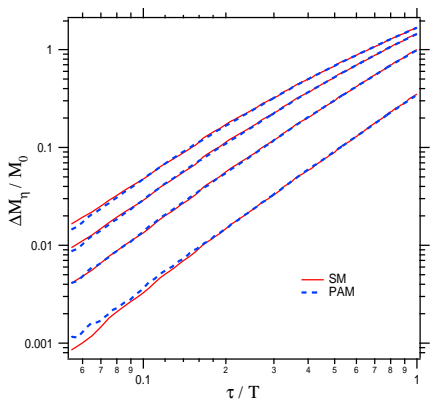}
\caption{(Color online)  \label{fig4}  
Log-log plots of the scaled separation $\Delta M_\eta(T,\tau)/M_0(T)$  
as a function of scaled time $\tau/T$ for several scaled perturbation 
strength $\eta/\eta_{th}=0.1,0.2,0.3,0.4$ from below at reversal time $T=400$.
 The data are for SM with $K=12$ and $\hbar=\frac{2\pi 975}{2^{20}}$
and PAM with $M=3,\epsilon=0.5$. 
}
\end{center}
\end{figure}
Such a universal behavior can be explained if we admit
the universality of the time-reversal characteristics $\irrev$ 
demonstrated by Fig.3, namely
\begin{eqnarray}
\label{scale1}
     \irrev = F(\frac{\eta}{\etath(T)}). 
\end{eqnarray}
Next we suppose the stationarity of the time-reversed 
dynamics, which means that for the same $\eta$ the 
difference $\Delta M_\eta \equiv M_\eta(T+\tau)-M_0(T+\tau)$ 
does not depend on the reversal time $T$, i.e., 
\begin{eqnarray}
\label{stationary}
\Delta M_\eta(T,\tau)=G(\eta,\tau), 
\end{eqnarray}
where $G$ is a function depending only on $\eta$ and $\tau$.
We do not give explicit evidences for the stationarity 
hypothesis, however, extended numerical examination supports 
the validity of this plausible hypothesis. Equations (\ref{scale1})
and (\ref{stationary}) claim that 
$G(\eta,\tau)=D\tau F(\frac{\eta}{\etath(\tau)})$, which is immediately followed
by the relation 
\begin{eqnarray}
\label{scale2}
\irrev &=& \frac{\tau}{T}F(\frac{\eta}{\etath(T)}\{\frac{\tau}{T}\}^{\chi}), 
\end{eqnarray}
where $\etath(T) \propto T^{-\chi}$. 
The index $\chi$ is determined by the type of perturbation as,
$\chi=1/2$ for perpendicular $\eta-$shift and 
$\chi=0$ for parallel $\eta-$shift. 
Thus, $\Delta M_\eta(T,\tau)/M_0(T)$ is determined only by 
the scaled perturbation strength $\eta/\etath$ and the scaled 
time $\tau/T$ for the reversal time $T(>T_{th})$.  

{\it In conclusion:}
The scaled universal relation (\ref{scale1}) implies $\irrev=2$ in 
the post quantum region, which means the complete loss of initial 
memory in that region, but in the quantum region it reaches zero in 
a common way independent of the reversal time $T$. 
The latter feature reveals an excellent stability and 
memory effect of the quantum systems against the perturbation in the 
quantum region, which can never be expected for their classical
counterparts.
On the contrary, the stationarity of
the system means that the time-reversed dynamics is free from
the past memory at the very time at which the time-reversed 
operation is applied. These two apparently contradictory features 
are combined to yield the universal scaled time-reversed dynamics 
in quantum normal diffusion.

Quantum characteristics of time-reversibility has been explored using 
four kinds of quantum maps exhibiting well-behaved normal diffusion.
In spite of the quite different nature of four models,
loss of initial memory is controlled by an universal quantum parameter 
called the least quantum perturbation unit(LQPU), and the time-reversed
dynamics as well as time-reversal characteristics is universal 
independent of the details of quantum maps.

We conjecture that the universal time irreversible characteristics 
demonstrated here is commonly shared by the ``most irreversible'' class of
quantum systems which require the most delicate control in order 
to retrieve the initial memory.
It should be stressed that such a universal feature is observed 
only for normal diffusion
systems, and the time-reversibility of localizing system and delocalized 
system without normal diffusion deviates much from the universal behavior 
in an non-universal way \cite{yamada10}. 
The quantum systems exhibiting the normal diffusion would be 
the most unstable class of simple 
quantum systems which models "quantum irreversibility".
Moreover, to clarify the quantum irreversibility would provide basic 
knowledge for the ultimate origin of intrinsic dissipation of small 
quantum systems.

This work is partly supported by Japanese people's tax via MEXT,
and the authors would like to acknowledge them.
They are also very grateful to Dr. T.Tsuji and and Koike memorial
house for using the facilities during this study.




\begin{thebibliography}{00}
\bibitem{footnote}
The classical SSM is exponentially unstable irrespective of 
the magnitude of $\epsilon$.
\bibitem{footnote2}
In the case of  
the parallel perturbation shifting is $\eta_{th}=\hbar$
because $\Delta x=2\pi$.
\end{thebibliography}
\end{document}